\documentclass[twocolumn,tighten]{aastex62}

\graphicspath{{./}{figures/}}

\shorttitle{Storms on Hot-Jupiters}
\shortauthors{Cho, Skinner, \& Thrastarson}

\begin{document}

\title{Storms, Variability, and Multiple Equilibria on Hot-Jupiters}

\correspondingauthor{James Y-K. Cho}
\email{jcho@flatironinstitute.org}

\author{James Y-K. Cho}

\affil{CCA, Flatiron Institute, New York, NY 10010, USA}

\author{Jack W. Skinner}

\affil{School of Physics and Astronomy, Queen Mary University of London, London
  E1 4NS, UK}

\author{Heidar Th. Thrastarson}

\affil{Jet Propulsion Laboratory, California Institute of Technology, Pasadena,
  CA 91109, USA}

\begin{abstract}

Observations of hot-Jupiter atmospheres show large variations in the location of
the ``hot spot'' and the amplitude of spectral features.  Atmospheric flow
simulations using the commonly employed forcing and initialization have
generally produced a large, monolithic patch of stationary hot area located
eastward of the substellar point at $\sim$\,$3\!\times\!  10^{-3}$\,MPa pressure
level.  Here we perform high-resolution (up to T682) pseudospectral simulations
that accurately capture small-scale eddies and waves, inherent in hot-Jupiter
atmospheres due to ageostrophy.  The atmospheres contain a large number of
intense storms over a wide range of scales, including the planetary-scale.  The
latter sized storms dictate the large-scale spatial distribution and temporal
variability of hot, as well as cold, regions over the planet.  In addition, the
large storms exhibit quasi-periodic life cycles within multiple equilibrium
states---all identifiable in the disk-integrated time series of the temperature
flux.

\end{abstract}


\section{Introduction} \label{sec:intro}

Hot-Jupiters orbit very close to their host stars.  Therefore, they are expected
to be in a 1:1 spin--orbit synchronized state and possess a high likelihood of
transiting their host stars.  Consequently, hot-Jupiter atmospheres are hitherto
the best observed of all the extrasolar planet atmospheres
\citep[e.g.,][]{Knutetal07,Swaietal08,Griletal08,Crosetal10,Armsetal16,Jacketal19,
  Belletal19,Esseetal19}.  Such observations represent the first step toward
assessing weather and climate on extrasolar planets.  However, to reliably
interpret and to optimally plan observations, accurate knowledge of the
three-dimensional (3D), global atmospheric flow and temperature patterns is
crucial.  A major reason for this is because dynamics forms the backbone for
accurate modeling of all the other important atmospheric processes (e.g.,
radiative transfer, clouds, photochemistry, and ionization).

Objects that are 1:1 spin--orbit synchronized are heated only on one side, the
``dayside.''  Such thermal forcing leads to global flow and temperature patterns
that are markedly different than those of the solar system planets.  Thus far,
simulations have either lacked the required horizontal resolution or have been
barotropic---i.e., two-dimensional \citep[2D;
  e.g.,][]{ShowGuil02,Choetal03,Choetal08,Showetal09,RausMen10,ThraCho10,
  Hengetal11,DobbAgol13,LiuShow13,Maynetal14,Polietal14,Choetal15,
  Mendetal16,22,23,Meno20}.  Hence, they have not been able to accurately
capture the dynamics of small-scales, important for the nonlinear interactions
with large-scales, and/or the crucial vertical coupling and variations
\citep{Choetal03,WatkCho10,ThraCho11,PoliCho12,Choetal15,SkinCho21a,
  SkinCho21b}.  In this Letter, we report on the results from a series of
high-resolution, 3D simulations using a highly accurate and well tested
pseudospectral code for extrasolar planets
\citep{Polietal14,Choetal15,SkinCho21a}.  The code solves the traditional
primitive equations \citep[e.g.,][]{Polietal14} with a high-order hyperviscosity
\citep[e.g.,][]{ChoPol96}.

\section{Setup} \label{sec:setup}

We employ an idealized setup that is commonly used in extrasolar planet
atmosphere modeling to generate the flow and temperature distributions starting
from an initial resting state \citep[e.g.,][]{LiuShow13,Choetal15}; see also
\citet{SkinCho21a,SkinCho21b} for all the physical and numerical parameters and
values belonging to the simulations presented in this Letter.  The setup
consists of ``relaxing'' the temperature field to a prescribed ``equilibrium
temperature'' distribution on a specified timescale at different pressure
levels.  Although highly idealized, this is a reasonable and practical first
representation of the thermal forcing in the absence of detailed
information---if the temperature perturbations from the equilibrium distribution
are not too large \citep{Choetal08}.  However, hot extrasolar planet atmospheres
can possess large temperature perturbations and are typically in a highly
ageostrophic regime, out of pressure gradient and Coriolis acceleration balance
\citep{Choetal15}.  This is because the rotation period~$\tau$ of the planet is
generally not short and the gravity wave speed~$c$ in the atmosphere is very
fast (e.g., $\tau \approx 3.025\!\times\!  10^5$~s and $c \sim 2.7\!\times\!
10^3$~m\,s$^{-1}$ for the planet described in this Letter).  In this situation,
the setup requires a very high resolution and stability for accurate
simulations.

\section{Results} \label{sec:results}

\begin{figure}
  \centerline{\includegraphics[scale=.085]{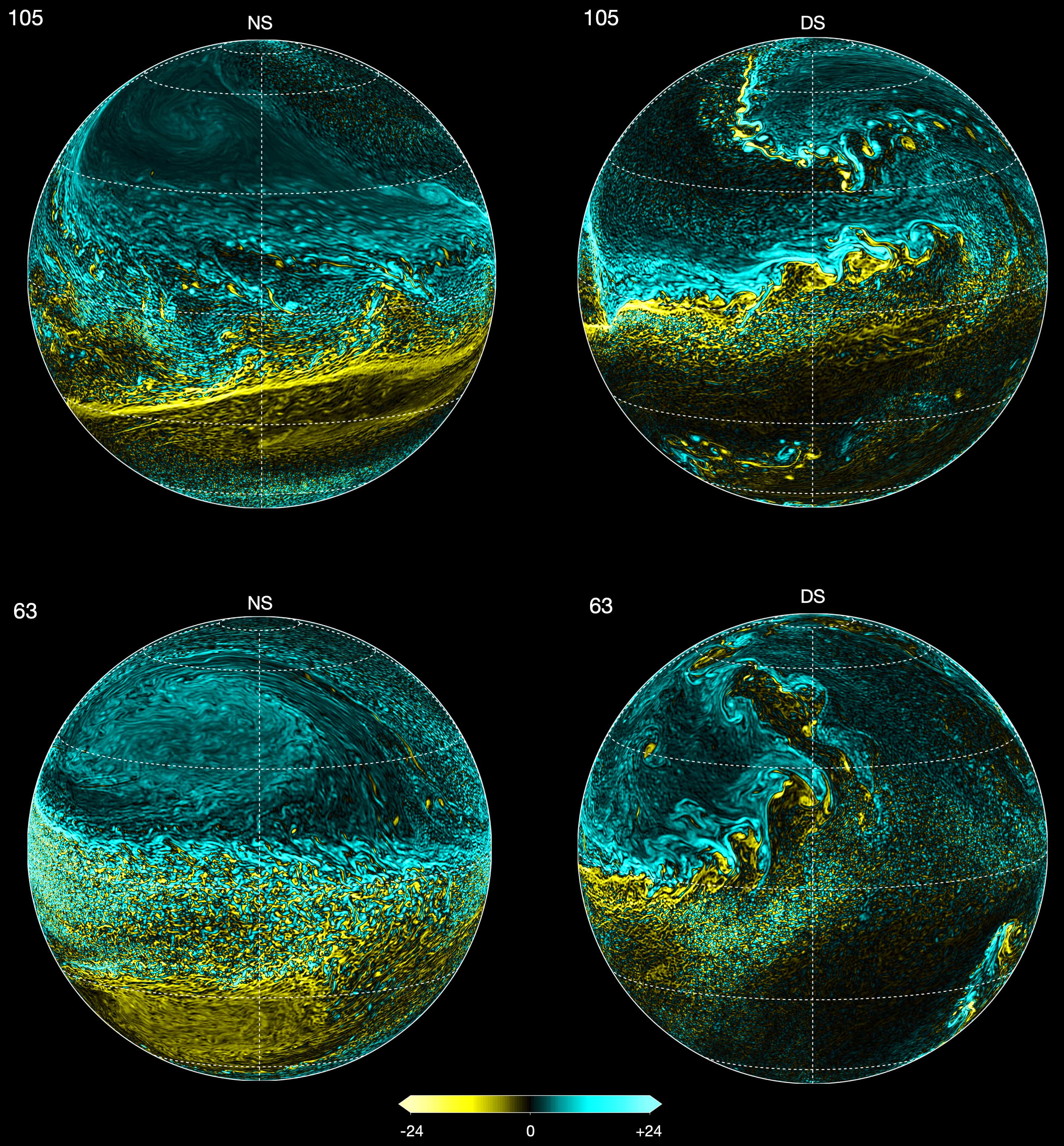}}
  \caption{Relative vorticity field $\zeta$, showing intense storms and
    meandering jet streams. The pressure levels are $0.00$5~MPa (top row) and
    $0.095$~MPa (bottom row); time (in the unit of a planetary rotation $\tau$)
    is indicated at the upper left of each frame.  The nightside (NS) and
    dayside~(DS) fields are shown centered on the antistellar and substellar
    points, respectively.  Cyan and yellow colors correspond to regions of
    $\zeta > 0$ and $\zeta < 0$, respectively, in the unit of $2\pi/\tau$.
    Highly dynamic storms form over a wide range of scales.  The planetary-scale
    storms exhibit quasi-periodic life cycles.}
  \label{fig:fig1}
\end{figure}

Figure~\ref{fig:fig1} shows the flow field (relative vorticity $\zeta$) from a
T682L20 resolution simulation with $\nabla^{16}$ hyperviscosity.  Here
``T682L20'' refers to 682 total wavenumbers and 682 zonal wavenumbers in the
spherical harmonics for each of the 20 pressure levels of the computational
domain \citep[][]{SkinCho21a}.  With the above viscosity, the resolution
corresponds to effectively at least an order of magnitude higher resolution in
the horizontal direction than those of past 3D simulations with comparable
vertical resolution \citep[e.g.,][]{Meno20}.  In other simulations discussed in
this work (T341L200), the vertical resolution is also two orders of magnitude
higher than those of past 2D hot-Jupiter simulations with comparable horizontal
resolution \citep[e.g,][]{Choetal03}.  The fields from near the top and bottom
of the T682L20 simulation are shown.\footnote[4]{The domain of this simulation
  extends down to 0.1\,MPa.  The 0.1\,MPa level is traditionally where the
  radius of a giant planet $R_p$ ($= 10^8$\,m, for the planet of this Letter) is
  measured and where most of the visible irradiation is expected to be fully
  absorbed on a hot-Jupiter \citep[e.g.,][]{Seagetal05}.  The proper location of
  the bottom (or the top) for simulations is currently unknown
  \citep{Choetal08}.}  Time $t$ is in the unit of $\tau$.

Several features are immediately apparent.  First, the flow is extremely
dynamic, and this characteristic persists over the entire duration of the
simulation (300\,$\tau$ here and 2000\,$\tau$ in the T341L200 simulations).  The
dynamism here is crucial because it actively redistributes temperature as well
as radiatively and chemically important species across the face of the planet.
Such spatiotemporal variability caused by evolving storms on a close-in planet
was first shown by \citet{Choetal03} in their 2D simulations employing T341
resolution with $\nabla^{16}$ hyperviscosity.  Second, the flow contains a very
large number of intense storms over the full range of scales---from the
planetary-scale down to nearly the dissipation-scale (near the scale of the
truncation wavenumber in the spherical harmonics).  At the planetary-scale,
there are two modons\footnote[5]{A modon is a long-lived, coherent pair of
  storms (a vortex couple) with opposite signs of $\zeta$ \citep{Ster75}.}: one
comprising a pair of cyclones---e.g., at $t = 63$ nightside (NS)---and the other
a (generally weaker) pair of anticyclones.\footnote[6]{Cyclones (anticyclones)
  are vortical structures that spin in the same (opposite) sense as the planet's
  north direction.}  Third, concurrent with the modons are sharp fronts and
high-speed jets that break and continuously spawn medium- and small-scale
storms.  The modons also directly generate energetic, small-scale gravity waves
\citep{WatkCho10} and storms as they attempt to adjust in the ageostrophic
environment \citep{LahaZeit12}.  Finally, the modons are also important in
blocking equatorial jets, breaking the zonal symmetry suggested in many past
simulations \citep[e.g.,][]{ShowPolv11,LiuShow13,Meno20}.  These features---as
well as others discussed below---are independent of the location of the bottom
of the domain (from 0.1\,MPa to 20\,MPa), provided the vertical range modeled is
well resolved with an adequate number of levels.

\begin{figure}
  \centerline{\includegraphics[scale=.084]{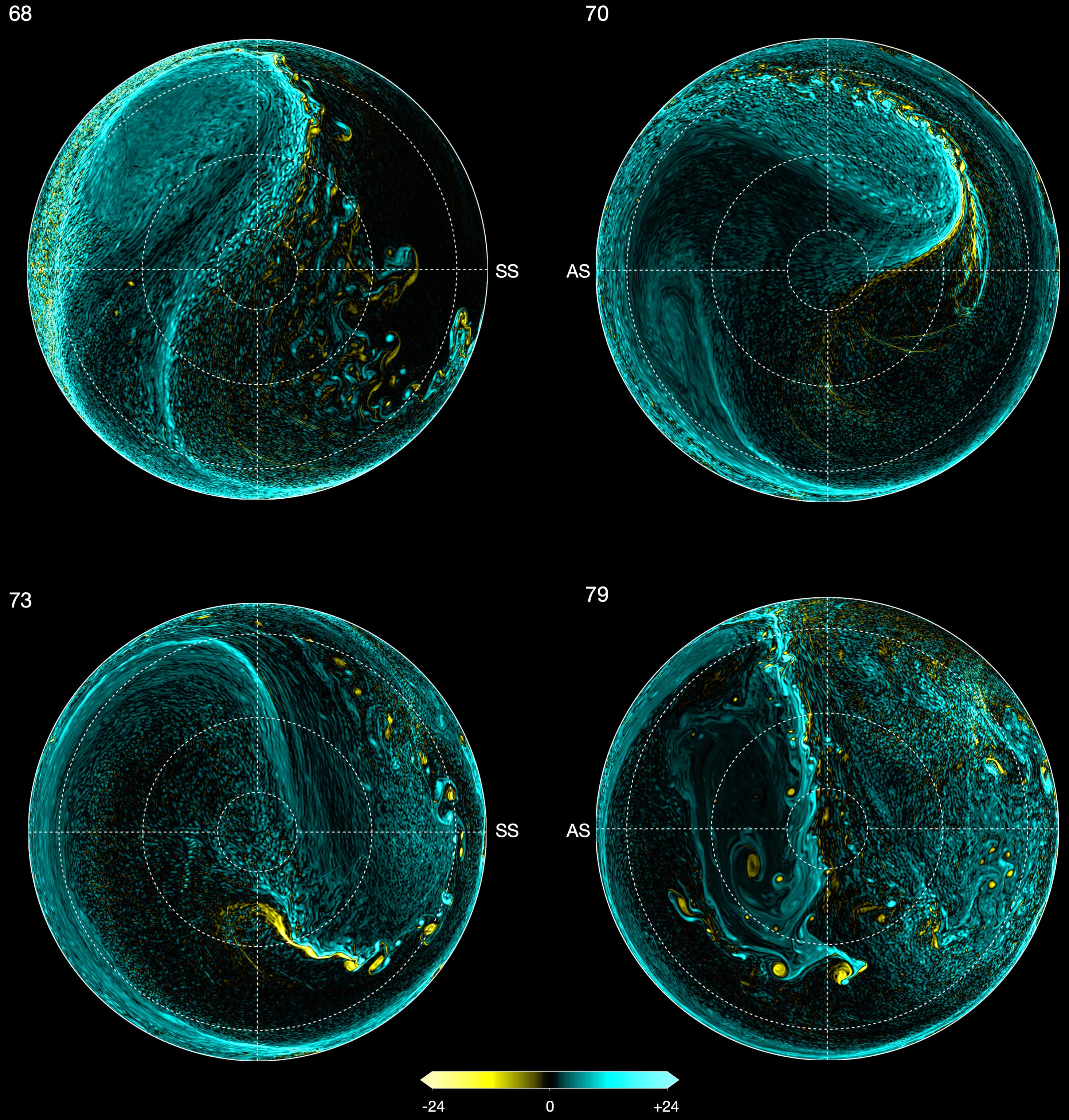}}
  \caption{The twilight phase of a cyclonic modon, in which the modon is
    ultimately replaced by a giant polar vortex; long-lived, medium-scale storms
    are also generated and dispersed widely across the planet.  The $\zeta$
    field at the $0.005$~MPa level from the simulation of Figure~\ref{fig:fig1}
    is shown, viewed from the north.  In each frame, the substellar point~(SS),
    antistellar point~(AS), eastern terminator, and western terminator are at
    the right, left, top, and bottom, respectively.  Such phase occurs in both
    the northern and southern hemispheres.}
  \label{fig:fig2}
\end{figure}

Medium- and small-scale storms form and move across hot-Jupiter atmospheres by
many different mechanisms
\citep{Choetal03,Choetal08,ThraCho10,WatkCho10,PoliCho12,Tsaietal14,Choetal15,
  Frometal16}.  These storms are important because they assist planetary-scale
storms in chaotically mixing the atmosphere on the global scale.
Figure~\ref{fig:fig2} shows one prominent, {\it recurring} mechanism---nonlinear
breaking and advection by a modon.  At $t = 68$, a large cyclonic modon that
initially formed just to the west of the substellar (SS) point has traversed
across the nightside and reached the eastern terminator (ET; top of the frame).
Throughout the traversal, the boundary of the modon continuously breaks and
rolls up into many storms.  At $t = 70$, the modon has moved past the
terminator, to a higher latitude, generating sharp fronts ahead (in longitude)
and below (in latitude); at this point, the northern half of the modon also
begins to separate from its partner cyclone in the southern hemisphere.  Here
one can also see gravity waves ahead of, and above, the modon---as well as a
second cyclone forming downstream, near the antistellar (AS) point.  By $t =
73$, the modon has passed over the SS point and intense storms generated at the
periphery of the modon are dispersed widely across the dayside, from the low- to
mid-latitudes.  Thereafter, the modon dissipates and is replaced (in this cycle)
by a planetary-scale vortex of nearly uniform $\zeta$ near the pole, at $t =
79$; note also the ``next-generation'' cyclonic modon, brewing near the SS point
at this time.  The overall motion of the modon is chaotic and not smooth from
its inception, but it is quasi-periodic.

\begin{figure*}
  \centerline{\includegraphics[scale=.095]{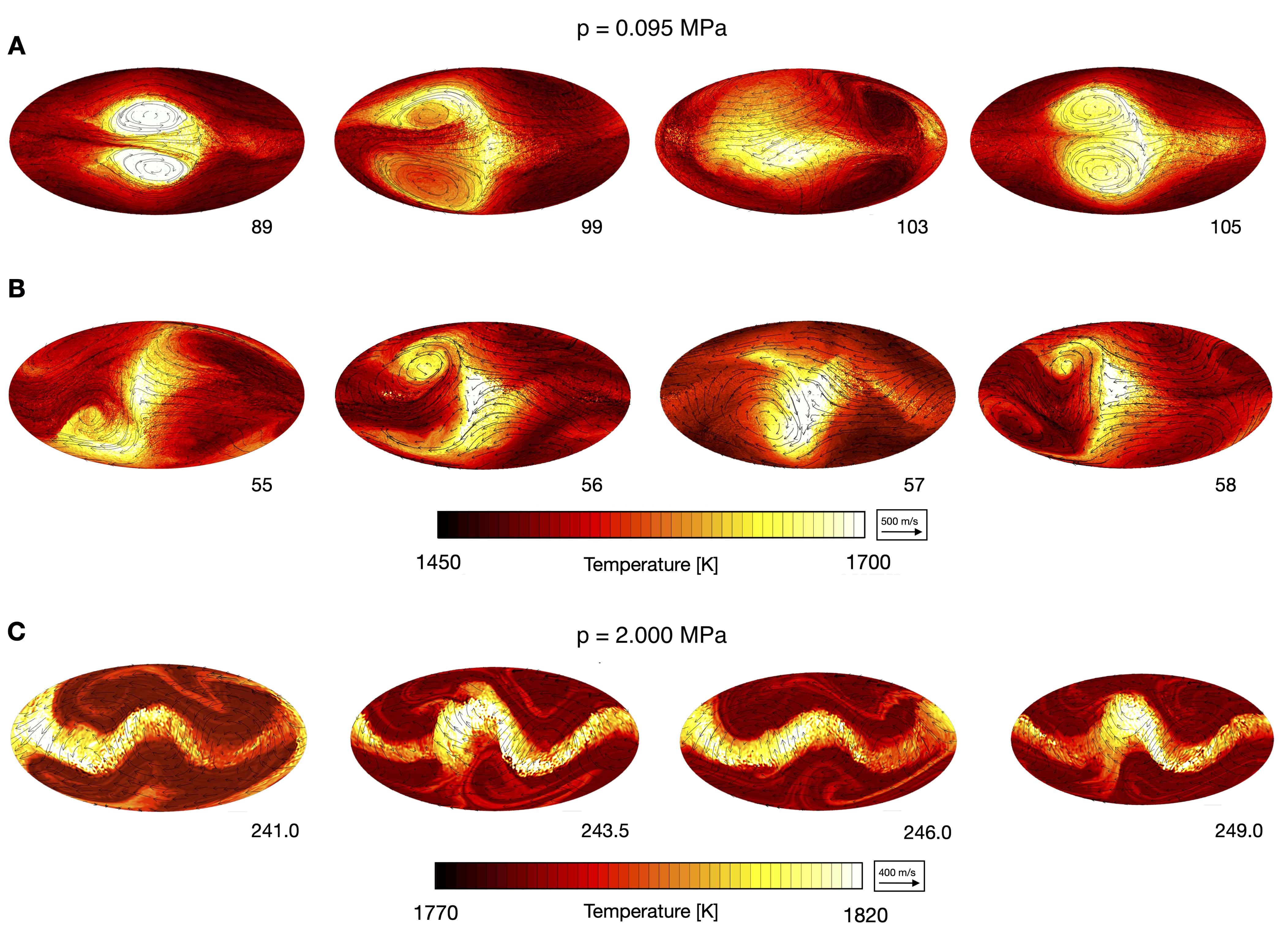}}
  \caption{Spatiotemporal variability in the temperature field $T$.  The fields
    at the $0.095$~MPa~(A and B) and $2.000$~MPa~(C) levels in Mollweide
    projection, centered at the SS point, are shown with time indicated at lower
    right of each frame.  Velocity vectors show the close association of the
    flow and $T$ fields.  The frames in (A) and (B) are from a T341L20
    simulation, and the frames in (C) are from a T341L200 simulation.
    Regardless of the domain range, different variability states exist at the
    same level (A and B) as well as at different levels (A and C). The
    variability is much more regular and wave-like in (C), compared with those
    in (A) and (B), but the thermal wave is highly nonlinear and periodically
    steepens, inducing elevated temperatures at different longitudes and
    latitudes at different times.  Modons initially form near the SS and AS
    points, sequester hot and cold masses of air, and then transport and mix the
    masses across the planet quasi-periodically as they move.}
  \label{fig:fig3}
\end{figure*}

In Figure~\ref{fig:fig3}, we illustrate several quasi-periodic patterns of the
temperature field $T$ induced by the modons.  There are more patterns than
presented.  The frames in Figures~\ref{fig:fig3}(A) and (B) are from a T341L20
simulation in which the pressure range of the domain is $[0, 0.1]$\,MPa.  The
frames in Figure~\ref{fig:fig3}(C) are from a T341L200 simulation in which the
pressure range of the domain is $[0, 10]$\,MPa.  The behaviors at the 0.095\,MPa
level shown are qualitatively similar in both of the simulations (as well as in
the T682L20 simulation above), albeit at different pressure levels.  The
T341L200 simulation is not level-wise converged with the T341L20 and T682L20
simulations because higher resolution (vertical and horizontal) is required and
the baroclinic structure of the flow is slightly different
\citep{SkinCho21a,SkinCho21b}.\footnote[7]{The flows in all three simulations
  are predominantly barotropic (i.e., vertically aligned).}  In the figure, the
velocity vectors are overlaid on the $T$ field and show the close relationship
between the flow and temperature.  In particular, modons sequester hot and cold
air masses and redistribute them over long distances.

For example, without the cyclonic modon (as well as other structures, such as
fronts, associated with strong flows), the $T$ field at $t = 89$ in
Figure~\ref{fig:fig3}(A) would be a simple, circular patch of ``hot spot''
centered at the SS point (instead of two disjointed patches); similarly, without
the anticyclonic modon, the $T$ field would be a single cold patch centered at
the AS point, rather than two separate cold patches at high latitudes.  As the
modons move (westward at $t = 99$), they transport large patches of hot and cold
air; both modons move and mix in both types of air. Note here that the hottest
region is well west of the SS point.  At $t = 103$, the anticyclonic modon (now
in the western hemisphere) has split apart, each half moving toward its
respective pole; here the anticyclones heat the polar regions as they move.
Simultaneously, the intense cyclonic modon just emerging from the nightside at
the ET advects cold air to the dayside from the nightside.  By this time, the
hottest area is again near the SS point, but the coldest area is not the AS
point.  At $t = 105$, the whole cycle has begun again, with the hottest region
$\sim$30$^\circ$ east of the SS point.  The period of this particular cycle is
$\sim$17\,$\tau$.

In contrast, two entirely different states are seen in
Figures~\ref{fig:fig3}\,(B) and (C).  In Figure~\ref{fig:fig3}(B), rather than
translating as a coherent structure, the cyclonic modon (in this cycle) is in a
``flapping state'': the northern and southern hemispheric halves alternately
``spin out''---transporting heat to higher latitudes in both hemispheres in a
periodic, sinuous manner.  The period of this cycle is $\sim\!3\,\tau$.  Note
that the two states in Figures~\ref{fig:fig3}(A) and (B) can switch back and
forth many times throughout the simulation.  In the deep part of the atmosphere
(Figure~\ref{fig:fig3}(C)), modons that initially formed at early times can
arrange themselves into a quartet of storms, often translating together westward
in a von K\'arm\'an vortex street--like configuration.  Here the temperature
variation and flow speed are comparatively smaller than those in the upper part
of the atmosphere, but they are no less dynamic---and, importantly, much more
periodic (with dominant periods of $\sim$$2.5\,\tau$ and $\sim$$5\,\tau$).  Due
to the slower speed in the deep region, the flow is quasi-geostrophic; hence,
hot and cold regions are generally very tightly associated with anticyclones and
cyclones, respectively.  Note also that thermal forcing is not applied at
pressure levels $\ge 1$\,MPa \citep[][]{LiuShow13,Choetal15}; hence, the
temperature variations are directly caused by the storms, which are entirely
powered by the much more vigorous activity at the lower pressure levels (Figure
1).

\begin{figure*}
  \centerline{\includegraphics[scale=0.095]{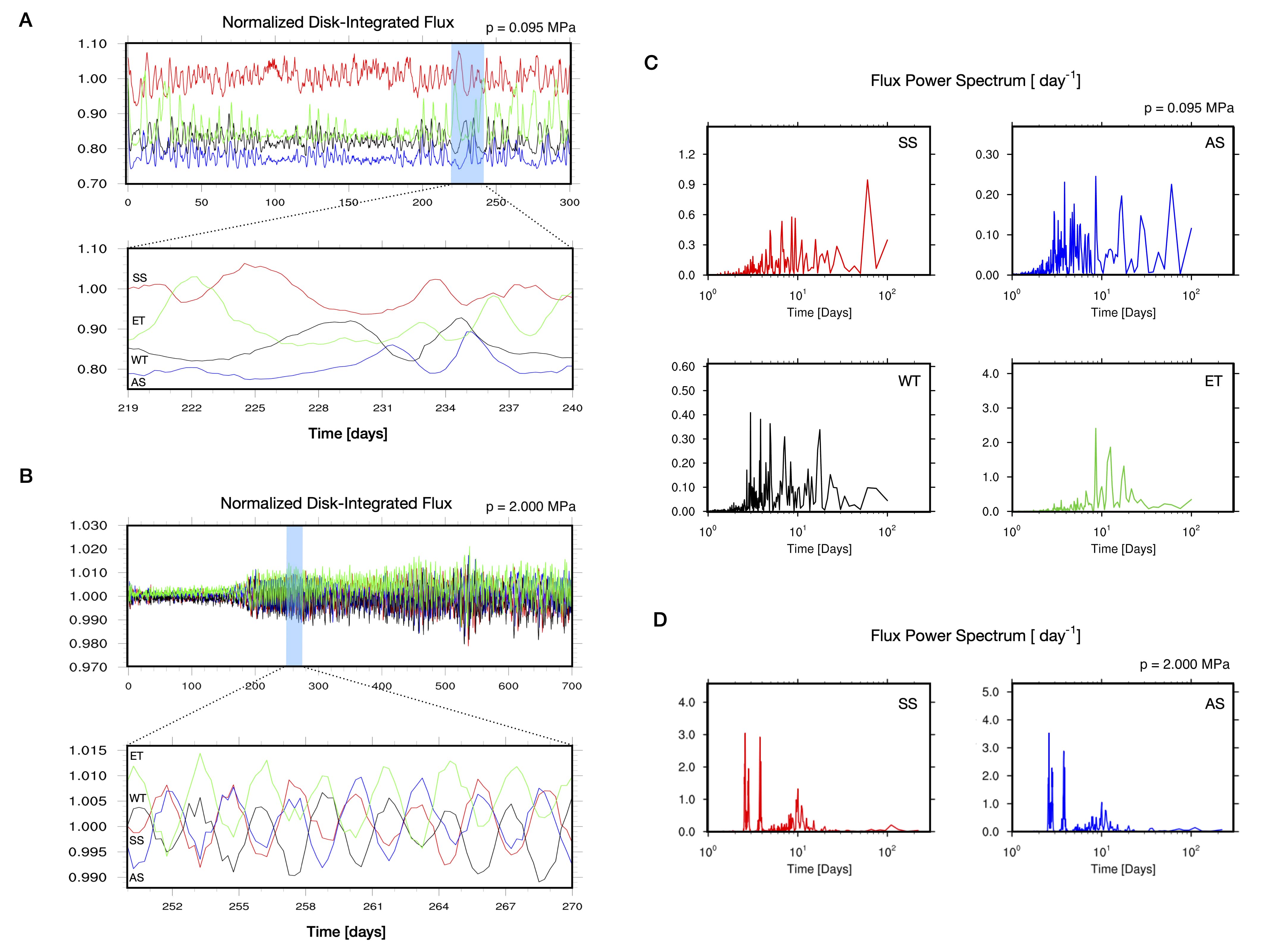}}
  \caption{ Time series (A, B) and power spectrum (C, D) of the disk-averaged
    temperature fluxes from the fields in Fig.~\ref{fig:fig3}. The averages are
    computed over disks centered at the substellar (SS), antistellar (AS),
    eastern terminator (ET), and western terminator (WT) longitudes at the
    equator and normalized by their initial values at the $0.095$\,MPa (A) and
    $2.000$\,MPa (B) pressure levels; the insets are magnifications of the
    periods shaded in blue.  Multiple states are seen at both levels.  In (A),
    the four fluxes can differ by $\sim$30\% among them, as well as within a
    single series; in (B), the variance and the amplitudes are much smaller than
    at the lower pressure level, but they jump to a new state with 5 times the
    old values at $t \approx 180$.  The hottest and coldest regions on the
    planet vary greatly in space and time at both levels shown.  The power
    spectra at the two levels are correspondingly different: the spectra in (C)
    are broad and densely peaked, while the spectra in (D) contain few dominant
    peaks, indicating a much more regular pattern. The spectra for WT and ET are
    essentially same as in (D) at $2.000$ MPa (not shown).
  }
  \label{fig:fig4}
\end{figure*}

All of these different states produce a distinct signature in the disk-averaged
temperature flux.  Figure~\ref{fig:fig4} shows the behavior of the atmosphere
over long durations.  Time series of the diskaveraged flux (proportional to
$T^{4}$ and adjusted for the surface normal orientation)\footnote[8]{Without
  radiative transfer, cloud, and other ingredients for additional physical
  realism, $T^4$ is an adequate measure of the equilibrium temperature flux.}
from the simulations presented in Figure~\ref{fig:fig3} are shown in
Figures~\ref{fig:fig4}(A) and (B), where a set of four time series are shown for
each pressure level.  The averages are obtained, centered at key points on the
planet: SS, AS, ET, and western terminator (WT).  Each set of temperature fluxes
are normalized by the initial mean value at the indicated pressure level; the
insets show the magnifications of the periods shaded in blue, illustrating the
clear in-phase and ``out-of-phase'' nature of the hot and cold ``spots'' over
the planet.  The period power spectra of the corresponding times series in
Figures~\ref{fig:fig4}(A) and (B) are shown in Figures~\ref{fig:fig4}(C) and
(D), respectively.  Note, the power spectra for WT and ET (not shown) are nearly
identical to those for SS and AS in Figure~\ref{fig:fig4}(D), the latter two of
which are themselves nearly identical to each other.  This is as expected from
the simple translation behavior in Figure~\ref{fig:fig3}C; in general, slight
variations in the spectra are observed as the quartet of storms transitions to
other configurations (not shown).

Such multiple states are readily seen at both pressure levels
(Figures~\ref{fig:fig4}(A) and (B)).  Although the states are generally
different at different levels, periodic states are present over long durations
at all levels.  However, the time series at higher pressure levels often undergo
state transitions correlated with transitions at lower pressure levels, usually
after a time delay---e.g., the transition to higher amplitude-variance state at
$t\!  \approx\! 180$ in Figure~\ref{fig:fig4}B ``kicking in'' following
energetic activity at Figure~\ref{fig:fig4}(A) (cf. ET series in
Figures~\ref{fig:fig4}(A) and (B), upper panels, starting at $t\! \approx \!
120$).  This helps to establish the quasi-barotropic structure over the whole
atmosphere.  We stress here that, despite the slower flow speed and smaller
temperature flux variation (compared to those at the lower pressure levels), the
flow {\it before} the transition is also quite active at the higher pressure
level; thus, temperature (and species) mixing occur at the higher pressure
level, even at this stage of the evolution when the fields appear nearly
``quiescent.''  Note that, at the higher pressure level, the temperature fluxes
can be very roughly divided into two groups according to their amplitudes
($\{\mbox{SS}, \mbox{AS}\}$ and $\{\mbox{ET}, \mbox{WT}\}$), in contrast to
those at the lower pressure level (Figures~\ref{fig:fig4}(A) and (B), lower
panels).  Hence, global temperature oscillations are not vertically (radially)
aligned.

Consistent with the time series, while the four power spectra at the lower
pressure level share common peaks, the spectra are all very distinct
(Figure~\ref{fig:fig4}(C)).  This is in marked contrast to the spectra at the
higher pressure level (Fig.~\ref{fig:fig4}(D)).  The spectra at the lower
pressure level are broad and densely peaked, but the spectra at the higher
pressure level essentially exhibit few peaks, with prominent ones located at
$t\! \sim\!  2.5$, $t\!  \sim\! 3.8$, and $t\! \sim\! 10$.  This is consistent
with the four series at the higher pressure level, which are essentially just
shifted in phase (Figure~\ref{fig:fig4}(B)).  Note also that, at this pressure
level, although there is a jump in temperature flux variance at $t\!  \approx\!
180$, the $\sim\! 2.5\,\tau$ period does not change before and after the jump.
However, period-shifts within a state do occur, in general, manifested as new
peaks in the spectra; for example, $\sim\! 5.5\,\tau$ and $\sim\! 8.5\,\tau$
periods before the jump gradually shift to $\sim\!  3.8\,\tau$ and $\sim\!
10\,\tau$ periods after the jump (Figure~\ref{fig:fig4}(D)), respectively, after
$t\!  \sim\! 300$.  Thus, some peaks remain constant across states, and some
peaks slowly change within a state.

\section{Discussion\label{sec:discussion}}

When the dynamics is adequately resolved over the required range of scales for
hot-Jupiters, our simulations show that intense storms induce variability on the
global scale---including causing hot and cold ``spots'' to be located both
eastward and westward of the SS and AS points, respectively, at different times.
Here we have used a setup (thermal forcing and initial-boundary condition) which
is commonly used in current extrasolar planet studies, in order to focus on
robust dynamics.  Quantitative aspects of the storms, variability, and states
may change depending on the precise setup and resolution used
\citep{ThraCho10,Choetal15}.  Hence, further investigations of the dependence on
the setup---at high resolution---should be carried out.  Based on the numerical
accuracy and convergence of the obtained solutions
\citep[][]{Polietal14,Choetal15,SkinCho21a,SkinCho21b}, the features reported
here are qualitatively robust and should apply generically to all 1:1
spin--orbit synchronized planets, including telluric ones.  Storms undergo
transitions to and from different persistent states or remain in one state over
a long duration, producing temperature flux signatures that may be observable.\\

We thank the Department of Astrophysical Sciences, Princeton University, where
some of this work was completed.  This research was supported in part by STFC
Consolidated grant 2017-2020 ST/P000592/1.  Part of this research was carried
out at the Jet Propulsion Laboratory, California Institute of Technology, under
a contract with the National Aeronautics and Space Administration.  We thank the
reviewer for helpful comments.

\end{document}